\documentclass[aps,prb,groupedaddress,twocolumn]{revtex4-1}

\usepackage{graphicx}
\usepackage{amsmath}
\usepackage{braket}
\usepackage{appendix}
\usepackage{bbm}

\begin{document}


\title{Influence of hopping selfenergy effects and quasiparticle degradation on the antiferromagnetic ordering in the bilayer honeycomb Hubbard model}


\author{Carsten Honerkamp}
\email[]{honerkamp@physik.rwth-aachen.de}
\affiliation{
Institute for Theoretical Solid State Physics, RWTH Aachen University,
D-52056 Aachen and JARA - Fundamentals of Future Information
Technology\\
}

\date{October 27, 2017}

\begin{abstract}
We study the Hubbard model on the AB-stacked bilayer honeycomb lattice with a repulsive onsite interaction $U$ in second order perturbation theory and in self-consistent random phase approximation.  We determine the changes in the antiferromagnetic magnetic ordering tendencies due to the real and imaginary parts of the selfenergy at the band crossing points. In particular we present an estimate for the threshold value $U^*$ below which the magnetic order is endangered by the splitting of the quadratic band touching points into four Dirac points by an interaction-induced interlayer skew hopping. For most of the parameter space however, the quasiparticle degradation by the frequency-dependence of the sublattice-diagonal selfenergies and the Dirac-cone steepening are more essential for suppressing the AF ordering tendencies considerably. Our results might help to understand to understand the energy scales obtained in renormalization group treatments of the same model and shed light on recent quantum Monte Carlo investigations about the fate of the magnetic ordering down to lower $U$.
\end{abstract}


\maketitle

\setlength{\parindent}{0pt}
\section{Introduction}
Hubbard models on single-layer and the AB-stacked bilayer honeycomb lattices have received attention in relation with graphene\cite{graphenerev} and as  relatively simple playgrounds for many-body physics in general (for some works, see Refs. \onlinecite{sorella1992,herbut2006,honerkamp2008,vafek,fanzhang,lang2012,schererbi,scherertri}. As distinctive feature compared to the single-layer honeycomb lattice, the simplest version of a tight-binding electronic  band structure on the AB-stacked bilayer with just nearest neighbor hopping $\gamma_0$ exhibits quadratic band crossing points, instead of Dirac points with linear dispersion\cite{mccann_rev}. This causes a nonzero density of states at the Fermi level of the half-filled band and hence leads to a stronger tendency towards interaction-induced instabilities at low temperatures. This is in contrast with the single layer, where a threshold interaction strength $U_c$ ($\sim 3.6\gamma_0$ for the onsite Hubbard interaction only according to quantum Monte Carlo (QMC)\cite{mengsorella}) has to be exceeded in order to destabilize the semi-metallic state in favor of gapped long-range ordered state. 

Until recently, it was generally believed that the interacting ground state of the half-filled spin-$\frac{1}{2}$ model with nearest-neighbor hopping and onsite repulsion only exhibits layered antiferromagnetic order for all, i.e. also arbitrarily weak repulsive onsite interactions $U$. This theoretical statement (see, e.g. Ref. \onlinecite{lang2012,schererbi}) was related with experimental reports about spectral gaps openings around the Fermi level at low temperatures\cite{laubi}.
The instability towards long-range order induced by electronic interactions is most easily studied in the Hubbard model with pure onsite interactions on the stacked honeycomb lattice. There, besides meanfield theory and renormalization group (RG) treatments, also rather controlled Quantum Monte Carlo studies are possible\cite{lang2012}. They give a quantitative account of the size of the ground state excitation gaps depending in the strength of the Hubbard repulsion $U$ and the interlayer hopping $\gamma_1$. In Ref. \onlinecite{lang2012}, these numbers were compared with those obtained by functional RG (fRG) and meanfield theory. It was found that for a given $U$ the QMC gaps are significantly smaller by almost one order of magnitude. This is not surprising as also in the single layer the critical $U_c$ for the onset of ordering is lower, $\sim 2.2\gamma_0$ in meanfield\cite{sorella1992} and $\sim 2.8\gamma_0$ in fRG\cite{sanchez}.  Of course, the latter two techniques are ultimately much less controlled than QMC, in particular because these  calculations were lacking selfenergy corrections to the electronic propagators. One might ask if one can understand the main factors that lead to these differences. The value of $U_c$ and the energy scales for ordering in the bilayer system may constitute valuable targets for methodical benchmarking. This constitutes one of the motivations for the present paper.

In addition to these quantitative issues, a recent QMC study\cite{lang2016} raised doubts on the simple picture of an ordering the bilayer for infinitesimally small interactions due to the nonzero density of states. These doubts were developed specifically for the Hubbard model  with pure onsite interactions on the bilayer, however with arguments that would also matter in the case of more general interactions and that would also have implications for the experiments that see gap openings in bilayer graphene. The main question that arose from the finite-size scaling of the QMC data is if the magnetic order persists down to lowest $U$. For values of $U \le 2.5t$, the system-size-extrapolated data for the magnetic ordered moments did not seem to converge to a nonzero value. The authors of Ref. \onlinecite{lang2016} then argued on general grounds that the interactions may generate symmetry-allowed linear terms in the dispersion that replace the quadratic two band crossing points of the simple 'plain-vanilla' model with two sets of four Dirac points. This in turn causes a nonzero threshold value for any interaction-induced ground state ordering, analogous to the single-layer case, where the critical value for the occurrence of a magnetically ordered ground state is $U_c \sim 3.6t$ measured in units of the in-plane hopping constant $t$. More precisely, for a quadratic band crossing point with nonzero density of states $\rho_0$, one expects a scaling of the energy gap due to antiferromagnetic order as $\Delta_{\mathrm{AF}} \sim e^{-1/(\rho_0U)}$. Also other physical quantities like the ground state energy reduction due to the ordering will be dominated by this energy scale. Compared to that, any interaction-induced hopping renormalization due to finite order processes would go $\sim U^2$. This causes a new small energy scale for a spectral modification $\sim U^4$ that will be larger than $\Delta_{\mathrm{AF}}$ as soon as $U$ is made small enough, smaller than a critical value $U^*$. Hence, the fate of the AF ordering at small $U<U^*$ is indeed questionable. 
 
Selfenergy corrections in graphene have been studied thus far mainly in field-theoretical context in the continuum model with long-range interactions. Early on it was pointed out\cite{gonzales,dassarma} that for the single layer model the Dirac points with Coulomb interactions exhibit marginal Fermi-liquid behavior and a substantial steepening of the velocity close to the Fermi level, in agreement with experimental findings\cite{elias}.  Furthermore, in the bilayer case, quasiparticle weight suppression and mass renormalization were included in a study by  Nandkishore and Levitov\cite{nandkishore}, again demonstrating non-Fermi liquid character. Barlas and Yang\cite{barlas} studied the quasiparticle scattering rate for Coulomb and onsite interactions, showing that in the bilayer the non-Fermi liquid properties exist already for short-range interactions. This was confirmed numerically for the Hubbard lattice model recently\cite{honerkampADP}. Linear-energy selfenergies in bilayer graphene were seen in photoemission experiments\cite{cheng}. What is missing so far is a quantitative study for the lattice model that explores the role of all possible selfenergy corrections on potential ordering instabilities.   
In this paper, we use second-order and RPA-summed infinite-order self-consistent perturbation theory to provide quantitative information about the size of selfenergy effects in mono- and bilayer honeycomb Hubbard models. This should also be a helpful exercise for later studies with more sophisticated methods like RG techniques. Besides studying quasiparticle degradation and band-stiffening effects, we estimate the size and impact of the interaction-generated $\gamma_3$. This clarifies the role of selfenergy corrections at least in the Hubbard model case and helps to interpret and connect various theoretical results for this situation. Furthermore our results will be useful information in the study of extended models that can be more directly related to experimental  graphene systems.  
 
 \section{Model and self-consistent perturbation theory}
 \subsection{Model}
Here we discuss the bilayer model defined on the AB-stacked honeycomb lattice. The single-layer model is obtained by simply taking the interlayer hopping equal to zero. We set the bond length, i.e. the distance between two nearest neighbor sites equal to unity and use the Bravais lattice vectors   
$
\vec{l}_1= 
\begin{pmatrix}
      \sqrt{3} \over 2   \\[1mm]
      \frac{3}{2}  
\end{pmatrix} $ and $ \qquad 
\vec{l}_2= 
\begin{pmatrix}
      -\sqrt{3} \over 2   \\[1mm]
      \frac{3}{2}  
\end{pmatrix} $.
The two vector components are for the $x$- and $y$-components in the lattice plane. The two layers are supposed to be stacked in the third direction. The bilayer unit cell then contains four sites, two per layer. We take the positions of these sites to be aligned along the $y$-direction. Sites 1 and 2 are 'on top of each other' in the origin of the unit cell, i.e. displaced by a vector in the $z$-direction, and connected by the interlayer hopping. Site 1 and 4 are nearest neighbors in one layer, displaced by a vector $\vec{n}= \vec{r}_2 - \vec{r}_1=(0,1/2)$. In the other layer, sites 2 and 3 in the same unit cell are displaced by $-\vec{n}= \vec{r}_3 - \vec{r}_2=(0,-1/2)$. The simplest 'plain-vanilla' bilayer model  has then the hopping Hamiltonian
\begin{equation}
\label{ }
\hat{H}_0 = \sum_{\vec{k} \in 1. \mathrm{BZ}, o,o' \atop s =\uparrow, \downarrow} 
h_{\alpha \beta} (\vec{k})  c_{\vec{k},s,\alpha}^\dagger  c_{\vec{k},s,\beta}    
\end{equation}
with the $4\times 4$ Hamiltonian matrix
\begin{equation}
\label{h0plain}
h_{\alpha\beta} (\vec{v})  = \left( \begin{array}{cccc}
0 & -\gamma_1 & 0 & h^*(\vec{k}) \\
-\gamma_1 & 0 & h(\vec{k}) & 0 \\
 0 & h^*(\vec{k}) & 0 & 0\\
 h (\vec{k}) &  0 & 0 & 0 \end{array} 
\right) \ ,
\end{equation}
 where the indices $\alpha$ and $\beta$ run over the four sites $1$ to $4$. The nearest neighbor hopping amplitude is 
 \begin{equation}
\label{ }
h (\vec{k}) = -\gamma_0 \, \left[ 1 +  \exp \left( i \vec{k}Ê\cdot \vec{l}_1 \right)  + \exp \left( i \vec{k}Ê\cdot \vec{l}_2 \right) \right]  \, .
\end{equation}
Note that we use the so-called 'proper gauge' with the Bravais lattice vectors in these exponentials. This maintains the reciprocal space periodicity. Regarding the parameters, we concentrate on the range of interlayer hopping $\gamma_1\le \gamma_0$. In graphene, realistic values would be $\gamma_1=0.1 \gamma_0$, but the intriguing QMC results\cite{lang2016} mentioned in the beginning were obtained mainly for $\gamma_1$ around $\gamma_0$ (in the correlated electrons community, the notation $t$ for the in-plane hopping $\gamma_0 $ is used, as well as $t_\perp$ for $\gamma_1$). 

The 'plain vanilla' Hamiltonian (\ref{h0plain}) can be considered a useful approximation to true bilayer graphene on an energy scale above a few meVs. It includes the marked difference of the AB-stacked bilayer in the electronic structure compared to the monolayer case. Instead of a Dirac point with a phase winding of the eigenvectors around it of $\pi$, the bilayer exhibits two quadratic band crossing points (QBCPs) with twice the phase winding.
Below an energy scale $\sim 1$meV, in real bilayer graphene additional terms become visible, as is nicely and quite throughly described in Ref. \onlinecite{mccann_rev}. Most notably, a skew hopping or 'trigonal warping' term $\gamma_3$ that connects sites 3 and 4 becomes noticeable that splits the two QBCPs into four Dirac points each. The energy scale below which the electronic structure is visibly modified is given by $\epsilon_L = \gamma_1 (\gamma_3/\gamma_0)^2/4$. Although in terms of pure values for BLG, $\gamma_3 \approx \gamma_1 \sim \gamma_0/10$, the square in $\epsilon_L$ reduces the effect of the skew hopping drastically.

The electronic interaction shall be idealized to pure onsite repulsion,
\begin{equation}
\label{ }
H_U = \sum_{i,\alpha} n_{i,\alpha,\uparrow} n_{i,\alpha,\downarrow} \, , 
\end{equation}
where the index $i$ runs over the $N$ bilayer unit cells and the sublattice label $\alpha$ over the four sites per bilayer unit cell.  Ab-initio theory\cite{wehling} can give estimates for the value of $U$. Coupling constants for further non-local interaction terms have also been computed and should be taken into account for a material-specific study. Here we ignore nonlocal interactions for the sake of simplicity, mainly because the perturbative calculations would become numerically more expensive.   

\subsection{Selfenergy in $U$ up to second order}
\begin{figure}
 \includegraphics[width=\columnwidth]{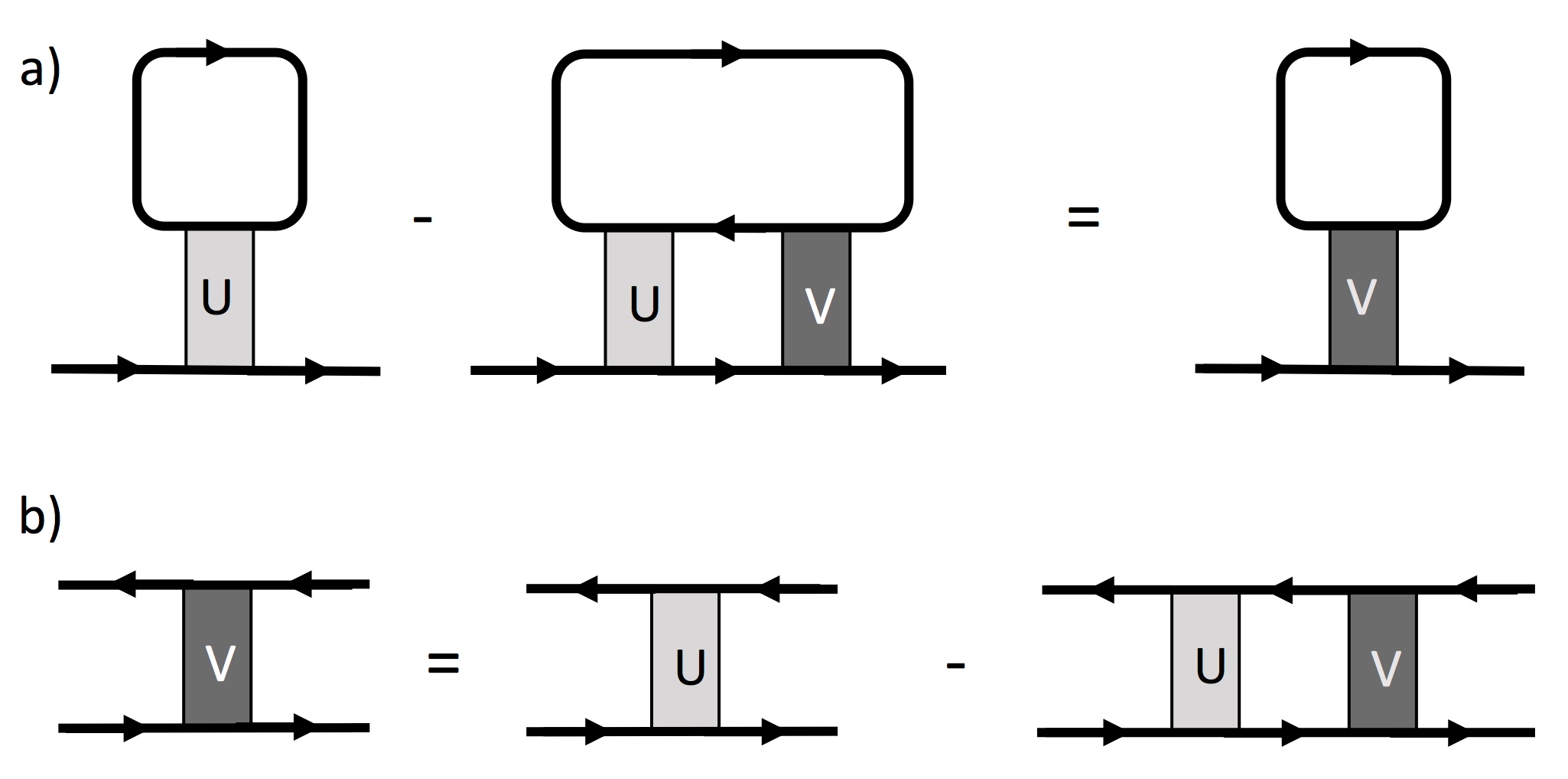}%
 \caption{ a) Left hand side: Schwinger-Dyson selfenergy. $U$ denotes the Hubbard interaction and $V$ in principle the fully renormalized one-particle irreducible vertex, here replaced by the RPA ladder summation of b). The internal lines are renormalized Green's functions. The lines below the interactions have spin index $s$, the upper lines have spin index $-s$. Right hand side: Effective one-loop expression for the selfenergy, when the ladder sum of b) is used. The relative minus signs come from the orders of the perturbation expansion in these diagrams. For the bubble summation for the vertex, an analogous argument can be made and both summations add in the one-loop selfenergy.}
  \label{diafig}
\end{figure}

The main object of interest in this paper is the selfenergy caused by the Hubbard interaction and its back-effect on the AF ordering tendency. 
A compact form to set up a self-consistent theory that can treat the AF ordering due to short-range interactions and quasiparticle degradation on equal footing is the Schwinger-Dyson equation\cite{wetterich} for the self-energy. Using a full Green's function $G_{s, \alpha \alpha} (k) $ and interactions that depend on wavevectors $\vec{k}$, Matsubara frequencies $ik_0$ combined to a three-index $k$, spin indices $s= \uparrow,\downarrow$ and sublattice indices $\alpha, \beta \in 1, \dots 4$ , this exact relation reads in our model
\begin{eqnarray}
\Sigma_{\uparrow, \alpha\beta}(k) &=& \delta_{\alpha \beta} \, U \frac{T}{N} \sum_{k'} G_{\downarrow, \alpha \alpha} (k') \nonumber  \\ \label{schwidy} &&  - \frac{UT ^2}{N^2} \sum_{k',k'' }  
G_{\uparrow, \alpha \mu'} (k')   G_{\downarrow, \alpha \mu''} (k'')   \\ &&  \cdot \, G_{\downarrow, \mu \alpha} (k''') V_{\mu' \mu'' \beta \mu}  (k',k'',k,k''') \nonumber  \,  , 
\end{eqnarray}
with $k'''=k'+k''-k$. A diagrammatic expression can be seen in Fig. \ref{diafig} on the left side of a).
This self-energy is a 4-by-4 matrix in sublattice space. As we study the regime above the magnetic symmetry breaking, the selfenergy is spin-independent,  $\Sigma_{\uparrow, \alpha\beta}(k) = \Sigma_{\downarrow, \alpha\beta}(k)$ and there are also no spin-offdiagonal contributions.
The first term on the right hand side of  (\ref{schwidy}) is of one-loop type. It just contains the bare onsite interaction, and hence only the Hartree term with opposite spin but identical sublattice index survives. For equal sublattice densities and without average spin polarization, this will just produce a wavevector-, frequency-, spin- and sublattice-independent constant that can be absorbed into a readjusted chemical potential. 
In the second term on the right hand side of (\ref{schwidy}), which is of two-loop type, on the right hand of the three full Green's functions one has the full one-particle irreducible (1PI) interaction vertex $V_{\mu' \mu'' \beta \mu}  (k',k'',k,k'+k''-k)$. This 1PI vertex is not known a priori. 
Our index convention is that the first two indices belong to the incoming lines, and the second two to the outgoing lines. 
 As the bare Hubbard interaction as the left vertex of the two-loop diagram mediates an interaction only between opposite spins, the spin indices of the first incoming line and the first outgoing line, i.e. $k'$ and $k$, have to be the same and the spin indices belonging to $k''$ and $k'+k''-k$ have to opposite. Hence there is no factor of two for a spin summation.

Of course one now has to employ an approximation for the 1PI vertex. The simplest approximation, which only mildly feels the AF ordering tendencies at finite $T$, is to use the bare Hubbard interaction, i.e. to write
\begin{equation}
\label{U1PI}
V^{(1)}_{\mu' \mu'' \beta \mu}  (k',k'',k,k'+k''-k) = U \delta_{\mu' \mu''} \delta_{\mu' \beta } \delta_{\beta \mu} \, .  
\end{equation} 

This leads to second-order perturbation theory if we use bare Green's functions on the internal lines of the two-loop diagram. The free Green's function can be expressed via the Lehmann representation as
\begin{equation}
\label{lehmann}
G^{(0)}_{s,\alpha\beta}(ik_0, \vec{k} ) = \sum_b \frac{u_{\alpha,b} (\vec{k}) u^*_{\beta b} (\vec{k}) }{i k_0 - \epsilon_b({\vec{k}}) } \, . 
\end{equation}
Here, $u_{\alpha,b} (\vec{k})$ denotes the respective components of the sublattice-to-bands transformation with band index $b=1, \dots, 4$ and energy bands $\epsilon_b (\vec{k})$. Below we will first use second-order perturbation theory to analyze the selfenergy components that are generated.  The full matrix Green's function can be obtained via Dyson's equation,
\begin{equation}
\label{ }
\hat{G}^{(0)}_{s}(ik_0, \vec{k} )  = \left[ i k_0 \mathbbm{1}  - \hat{h} (\vec{k}) - \hat{\Sigma}(k) \right]^{-1} \, 
\end{equation}
with the Hamiltonian matrix defined in (\ref{h0plain}). 

\subsection{Random phase approximation}
Here, we are mainly interested in effects close to the AF ordering instability. The latter can be implemented perturbativley in the random phase approximation (RPA), i.e. in an infinite-order summation of certain classes of diagrams that are functions of just one wavevector/frequency $q$. Diagrammatically, as shown in Fig. \ref{diafig}, it is rather straightforward to show that the two-loop contribution to the self-energy in Eq. \ref{schwidy} can be rewritten as
\begin{equation}
\label{RPAsig}
\Sigma^{2-\mathrm{loop}}_{\alpha\beta}(k) = \frac{T}{N} \sum_{q} G_{\alpha \beta} (k+q) V^{\mathrm{RPA}}_{\alpha \beta} (q) \, ,
\end{equation}
where the RPA-interaction consists of an infinite ladder-diagram sum (in some communities also known as T-matrix summation) and an infinite sum over an odd number (as we have to end up with the same spin again) of at least three bubble diagrams, 
\begin{equation}
\label{ }
 V^{\mathrm{RPA}}_{\alpha \beta} (q)  =  V^{\mathrm{RPA, ladder}}_{\alpha \beta} (q)  + V^{\mathrm{RPA, bubbles}}_{\alpha \beta} (q)  \, 
\end{equation}
with (using matrix notation)
\begin{equation}
\label{Vladder}
\hat{V}^{\mathrm{RPA, ladder}} (q) = U^2 \hat{L} (q) \left[ \mathbbm{1} + U \hat{L}  (q) \right]^{-1} \, , 
\end{equation}
and 
\begin{equation}
\label{ }
\hat{V}^{\mathrm{RPA, bubbles}} (q) = U^4 \hat{L}^3(q) \left[ \mathbbm{1} - U^2 \hat{L}^2  (q) \right]^{-1} \, .
\end{equation}
With these subsets of diagrams, the obtained 1PI effective interaction vertex still only depends on two sublattice indices $\alpha$, $\beta$ for two pairs of incoming and outgoing lines and has opposite spin projections for these two lines for the ladder sum and equal projections for the bubble sum (the latter difference does not play a role in our formulae, as the Green's function is spin-independent).  In these equations, the sublattice-resolved matrix-valued particle-hole bubble is 
\begin{equation}
\label{ }
{L}_{\alpha \beta} (q) = \frac{T}{N}Ê\sum_{k} G_{\alpha \beta} (k) G_{\beta \alpha} (k+q) \, . 
\end{equation}
With full Green's functions $G_{\alpha \beta} (k)$ on the internal lines of the bubble, we arrive at selfconsistent RPA. We can also define a 'bare' RPA$^{(0)}$ by leaving out the selfenergy on the internal lines.

The AF instability is seen as a divergence in the 1PI vertex or, more directly as a divergence in the RPA spin susceptibility
\begin{equation}
\label{chiaf}
\hat{\chi}_s^{\mathrm{RPA}} (q) =   \hat{L} (q) \left[ \mathbbm{1} + U \hat{L}  (q) \right]^{-1} \, 
\end{equation}
at $q=0$. For this to happen, in the square brackets, an eigenvalue of the matrix $\hat{L}  (q)$, has to approach $1/U$ when $U$ is increased or when $T$ is lowered. This is a straightforward generalization of the Stoner instabilities usually discussed in one-band models. We call the eigenvalue approaching $1/U$ in all what follows eig$_{\mathrm{AF}}$. The reason is that the eigenvector belonging to eig$_{\mathrm{AF}}$ is of staggered type, i.e. it changes its sign from one sublattice to the other within one plane and also along the bonds connecting the planes. Actually, in agreement with Ref. \onlinecite{lang2012} in the numerical evaluation of (\ref{chiaf}) for the lattice model, this eigenvector has larger components on the sublattice sites that are not connected by the interlayer hopping $\gamma_1$. This quantitative difference is mainly due to the fact that the low-energy bands that touch at $K/K'$ have larger weights on these sites. The 'denominator' $\left[ \mathbbm{1} + U \hat{L}  (q) \right]^{-1}$ can also be viewed as inverse Stoner factor which enhances the bare susceptibility $\hat{L}  (q) $. 

Obviously, near the AF instability $ U \hat{L}  (q) \approx - \mathbbm{1}$ (more strictly, this only holds in the subspace of the eigenvector whose eigenvalue approaches one). This can be used to argue that $V^{\mathrm{RPA, bubbles}}_{\alpha \beta}Ê (q) \approx V^{\mathrm{RPA, ladder}}_{\alpha \beta} (q) /2$, such that we can also estimate the  effective RPA interaction near the instability as 
\begin{equation}
\label{threehalves}
 V^{\mathrm{RPA}}_{\alpha \beta}Ê (q)  =  \frac{3}{2} \, V^{\mathrm{RPA, ladder}}_{\alpha \beta}Ê (q) \, .
\end{equation}
Hence the bubble contribution to the selfenergy mainly enhances the ladder term.
In the numerics described below we did runs with the factor $3/2$ and just with the ladder selfenergy, with qualitatively similar results. In what follows we show RPA data for the ladder RPA selfenergy only, in order to have a conservative view of the selfenergy impact. From RG treatments we know that the RPA ladder overestimates the AF ordering tendencies quantitatively, hence a more conservative treatment of the RPA selfenergy is possibly more realistic than using the factor $3/2$.

 \section{Second order selfenergy}
\begin{figure}
 \includegraphics[width=\columnwidth]{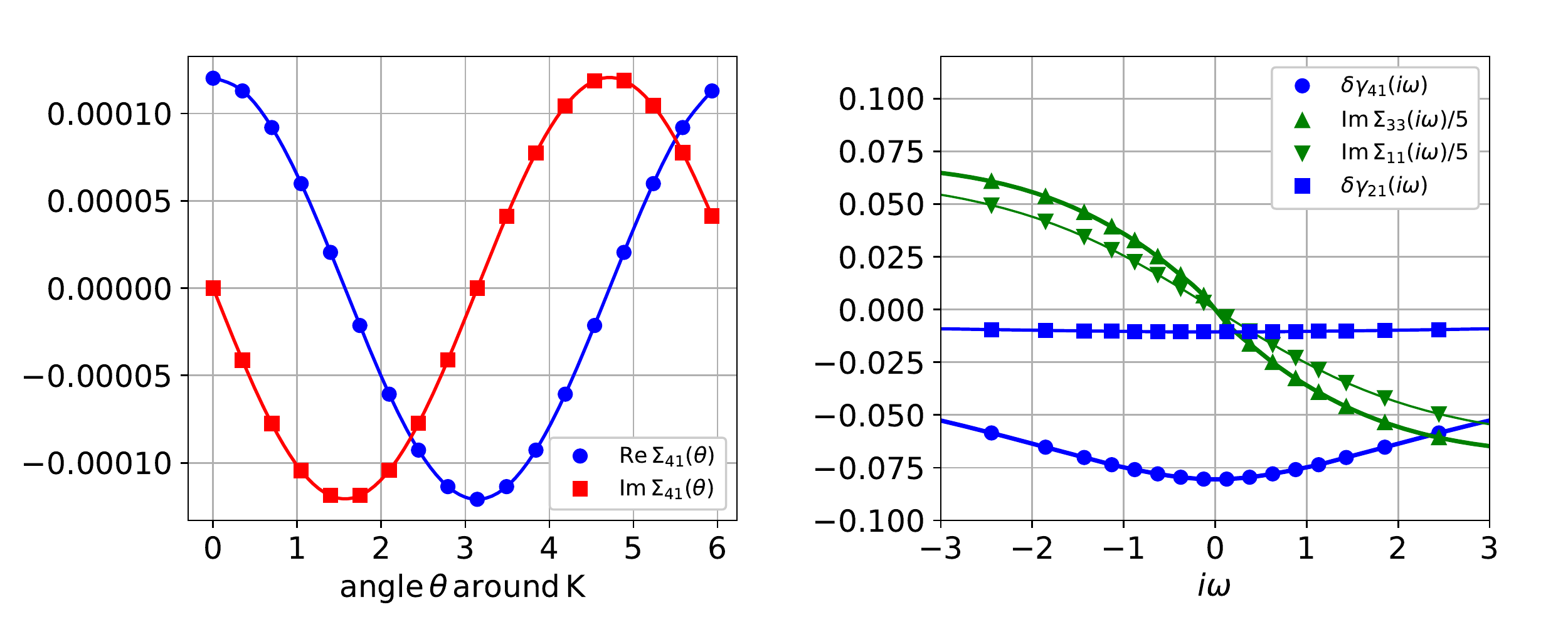}%
 \caption{Data for the selfenergy in 2nd order $U$, with $U=3.6\gamma_0$, $T=0.04\gamma_0$, $\gamma_1= 0.8\gamma_0$. Left panel: The data points (squares) show the angular variation of the real and imaginary parts of the intralayer intersite selfenergy $\Sigma_{41} (\vec{k} , i \omega = i\pi T) = \Sigma_{23} (\vec{k} , i \omega = \pi T) $ on a circle of radius $r=0.001$ in units of the inverse bond length around the $K$-point. The solid lines show the nearest neighbor hopping form factor, rescaled by a factor $\delta \gamma_{41}$ so as to match the data points. Right panel: Matsubara frequency dependence of the selfenergy parameters $\delta\gamma_{41} $, $\delta\gamma_{21} $ (interlayer hopping renormalization), Im $\Sigma_{11}=$Im $\Sigma_{22}$ and  Im $\Sigma_{33}=$ Im $\Sigma_{44}$ at $\vec{k}=K$, rescaled by a factor 5. The latter two functions determine the quasiparticle degradation in the bands. Orbitals 3 and 4 form the bands that touch at the Fermi level, while orbitals 1 and 2 constitute those bands that are gapped via the interlayer hopping $\gamma_1$. }
  \label{sig4133U2plot}
\end{figure}

\begin{figure}
 \includegraphics[width=\columnwidth]{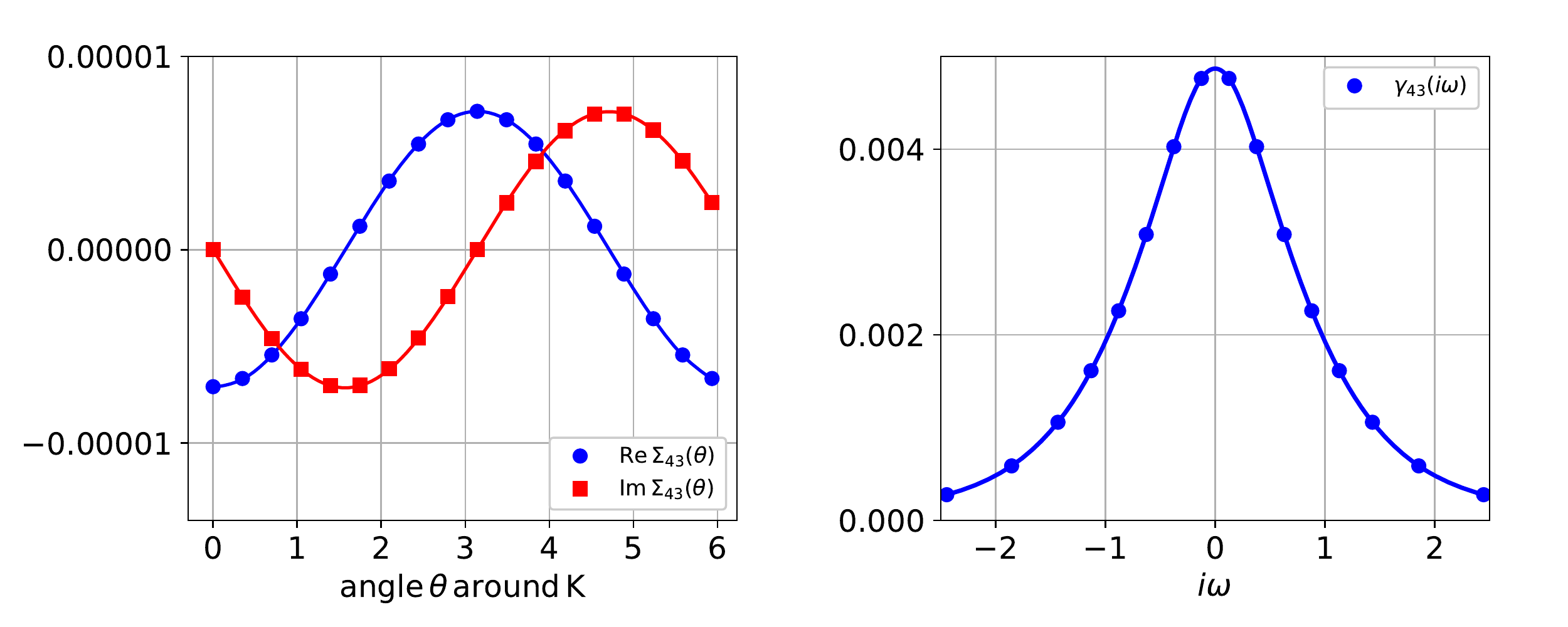}%
 \caption{ Data for the selfenergy in 2nd order $U$, with $U=3.6\gamma_0$, $T=0.04\gamma_0$, $\gamma_1= 0.8\gamma_0$. Left panel: The data points (squares) show the angular variation of the real and imaginary parts of the skew hopping interlayer selfenergy $\Sigma_{43} (\vec{k} , i \omega = i\pi T)  $ on a circle of radius $r=0.001$ in units of the inverse bond length around the $K$-point. The solid lines show the corresponding interlayer nearest-neighbor hopping form factor which is the complex conjugate of the normal intralayer hopping form factor, again rescaled by a factor $\delta \gamma_{43}$ so as to match the data points. Right panel: Matsubara frequency dependence of the skew hopping selfenergy parameter $\delta\gamma_{43} $. }
  \label{ga43U2plot}
\end{figure}

\begin{figure}
 \includegraphics[width=\columnwidth]{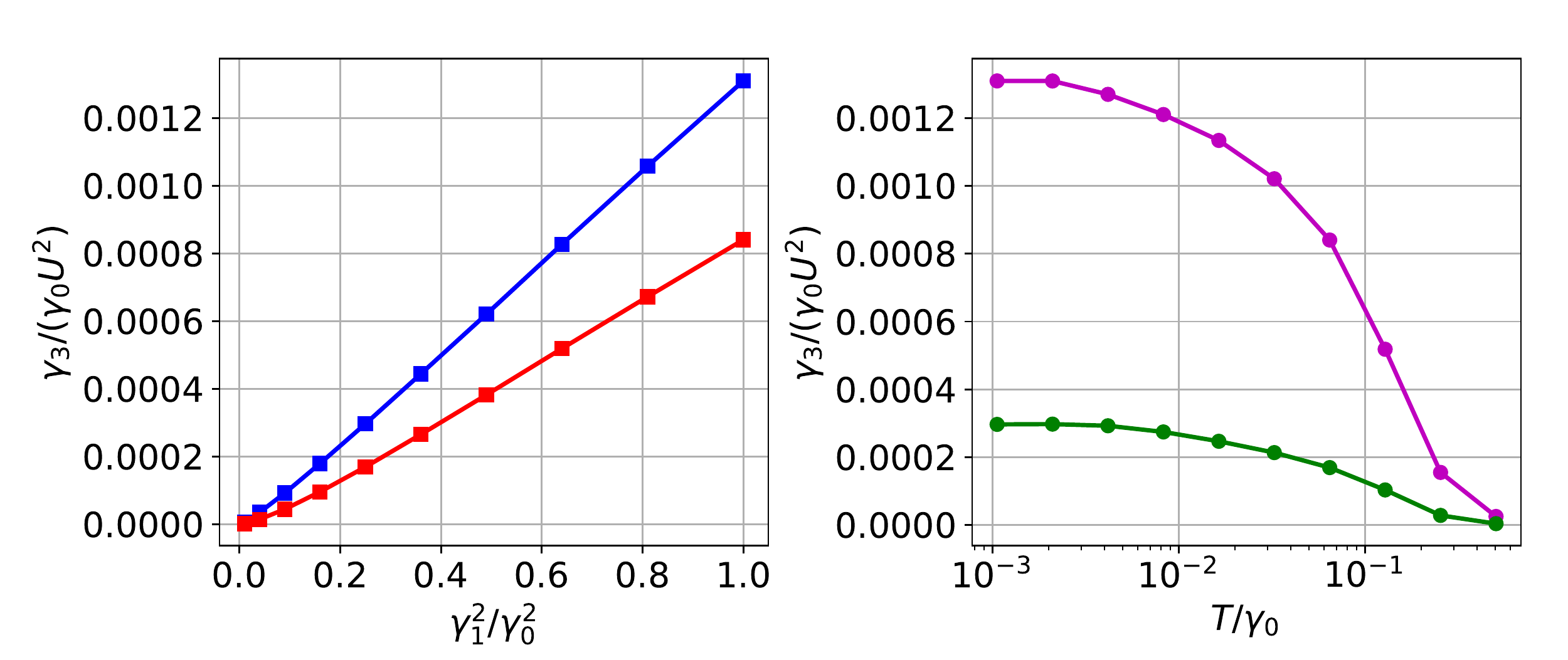}%
 \caption{Data for the selfenergy in 2nd order $U$. Left panel: Dependence of $\gamma_{43}(i\omega = i\pi T)$ (divided by $U^2$) versus interlayer hopping $\gamma_1$, which roughly follows $\gamma_1^2$, i.e. $\gamma_{43} \sim 0.001 \gamma_1^2 U^2/ \gamma_0^3$.  The upper line is for $T=0.001\gamma_0$, the lower one for $T=0.064\gamma_0$. Right panel: $T$-dependence of $\gamma_{43}/U^2$ at the lowest Matsubara frequency.  The upper line is for $\gamma_1=\gamma_0$, the lower line for $\gamma_1=0.5 \gamma_0$.}
  \label{Ga43TdepU2}
\end{figure}

For the start, let us look at the selfenergy matrix $\Sigma_{\alpha\beta} (\vec{k}, i \omega)$ in the sublattice basis on the Matsubara frequency axis and for wavevector $\vec{k}$ near the $K$, $K'$ points, obtained numerically in second oder perturbation theory in $U$.  We study the wavevector  dependence of $\Sigma_{\alpha\beta} (\vec{k}, i \omega)$ on four rings composed of 18 points each with radii 0.0001, 0.001, 0.01 and 0.1 (in units of the inverse bond length) around the $K$,$K'$ points. The different selfenergy matrix elements exhibit a quite simple dependence on the radius of these lines. They are either constant within a few percent or rise linearly. The radius-independent components have very little angular dependence while the linearly rising terms follow the nearest-neighbor form factors  $h(\vec{k})$ or $h^*(\vec{k})$ in the their angle-dependence. Here we list the various selfenergy terms with their main properties.
 \begin{itemize}
  \item The diagonal terms $\Sigma_{\alpha\alpha} (\vec{k}, i \omega)$ are basically wavevector-independent in the low-energy region considered with radius $\le 0.1$. The real parts are zero, in consistency with particle-hole symmetry. The imaginary parts show the typical frequency dependence of a weakly correlated itinerant system, with a linearly falling part form positive values at small negative frequencies to negative values at small positive frequencies. From the slope of this decrease, one can determine $Z$-factors via $Z_a=\left[1-i \partial_{i\omega}  \left.  \Sigma_{\alpha\alpha} (\vec{k}, i \omega) \right|_{i\omega=0}\right]^{-1}$. The selfenergy is larger and steeper in frequency on the two sublattice sites (called 3 and 4) that are not connected by the interlayer hopping, as these sites have larger amplitude in the bands that touch at the Fermi level.
  \item The intralayer hopping $\gamma_0$ between sites 1 and 4 or 2 and 3 gets renormalized to larger absolute values by elements like  $\Sigma_{14} (\vec{k}, i \omega)$. These elements follow in their real and imaginary parts very nicely the hopping form factor $h(\vec{k})$ (see left plot Figure \ref{sig4133U2plot}) and have a symmetric peak at small Matsubara frequencies around 0 (see right plot Figure \ref{sig4133U2plot}). From fitting to $h(\vec{k})$ one can extract a hopping increase $\delta \gamma_{41}$ of the order of 10$\%$ (for $U=3.6\gamma_0$) whose precise value  depends on how close the system is to a AF instability. It comes out negative and adds to the negative $-\gamma_0$ with $\gamma_0$ taken $>0$ in the bare dispersion. The effects of this is a Dirac cone steepening. Experimentally, such a velocity upward-renormalization is observed for suspended graphene and is understood as an effect of the long-range Coulomb interaction that peaks at small wavevectors\cite{elias}. In our case with a bare onsite Hubbard interaction, the particle-hole loops generate a $\vec{q}=0$-peak in the effective interaction that causes a similar effect on the hopping.
  \item The interlayer hopping $\gamma_1$ is also renormalized in second order in $U$. This comes from $\Sigma_{21} (\vec{k}, i \omega)$ , which turns out to be also peaked at small $|\omega|$ and negative. As the interlayer hopping is defined as $- \gamma_1$ in the Hamiltonian, with positive $\gamma_1$, the selfenergy terms acts to enhance the interlayer hopping. Overall this effect is quite mild, $\sim 1\%$ for $U=3.6\gamma_0$ and not strongly $T$-dependent.
  \item Next there is the interaction-generated interlayer skew hopping $\Sigma_{43} (\vec{k}, i \omega)$. Again the wavevector dependence of this selfenergy component is well described by the hopping form factor $h^*(\vec{k})$ (see left plot of Fig. \ref{ga43U2plot}). The need for a complex conjugation can be seen in the top view on the bilayer, where the three nearest neighboring "4"-sites of a a site "3" in the other layer are at inverted shift vectors compared to the nearest neighboring "2" that connect to "3" via inplane hopping.  The frequency-dependence of this term is again symmetric and peaks at small absolute values of the Matsubara frequency. Again, we can fit to the hopping form factor in order to extract a magnitude $\gamma_{43}$. This value comes out positive, i.e. opposite in sign to the intraplane hopping that we define to occur as $-\gamma_0$ in the Hamiltonian. It causes a split of the quadratic band crossing points at $K$, $K'$  into 4 Dirac points each, three of which are displaced along the BZ edges. In second order in $U$ we find roughly  $\gamma_{43} \sim 0.001 \gamma_1^2 U^2 / \gamma_0^3$. We are not sure if there is a simple argument to extract the quadratic dependence on the interlayer hopping $\gamma_1$, but the numerical data in the left plot of Fig. \ref{Ga43TdepU2} suggest such a behavior. The temperature dependence shows the that the term gets larger below $T\sim 0.1\gamma_0$ and saturates toward low $T$. 
\item There is also an interaction-generated term $\Sigma_{42} (\vec{k}, i \omega)$. It also follows the nearest-neighbor hopping angular dependence, but is odd in frequency. Its magnitude $\gamma_{42}(\pi T) $ is however of the order $10^{-4}\gamma_0$ such that we neglect this component in the further discussion. It is nevertheless kept in the self-consistent calculations presented later on. 
\end{itemize} 

Based on this data, we conclude that there are three different types of selfenergy effects that can be expected to work against the antiferromagnetic state that would be found in meanfield theory or in the random phase approximation without selfenergy inclusion. The rather strong Dirac cone steepening $\delta \gamma_0$ reduces the density of states $\rho_0 \sim \gamma_1/\gamma_0^2$ at low energies, also in presence of an interlayer coupling and a slightly increased $\gamma_1$ by the observed $\delta \gamma_1$. Second, the quasiparticle degradation due to the non-hermitian parts of the selfenergy, most notably by the imaginary parts of the diagonal elements, reduces the spectral weight at low frequencies.  This reduction can become noticeable, as the $Z$-factor of the bands at the Fermi level goes down with decreasing $T$ and reaches values below 0.65 for $U=3\gamma_0$. This indicates that the AF ordering transition might be affected by the quasiparticle degradation combined with the reduction of the density of states. We will see this more directly in the following in the self-consistent RPA treatment.

\section{Role of the interaction-induced skew hopping}
\begin{figure}
 \includegraphics[width=\columnwidth]{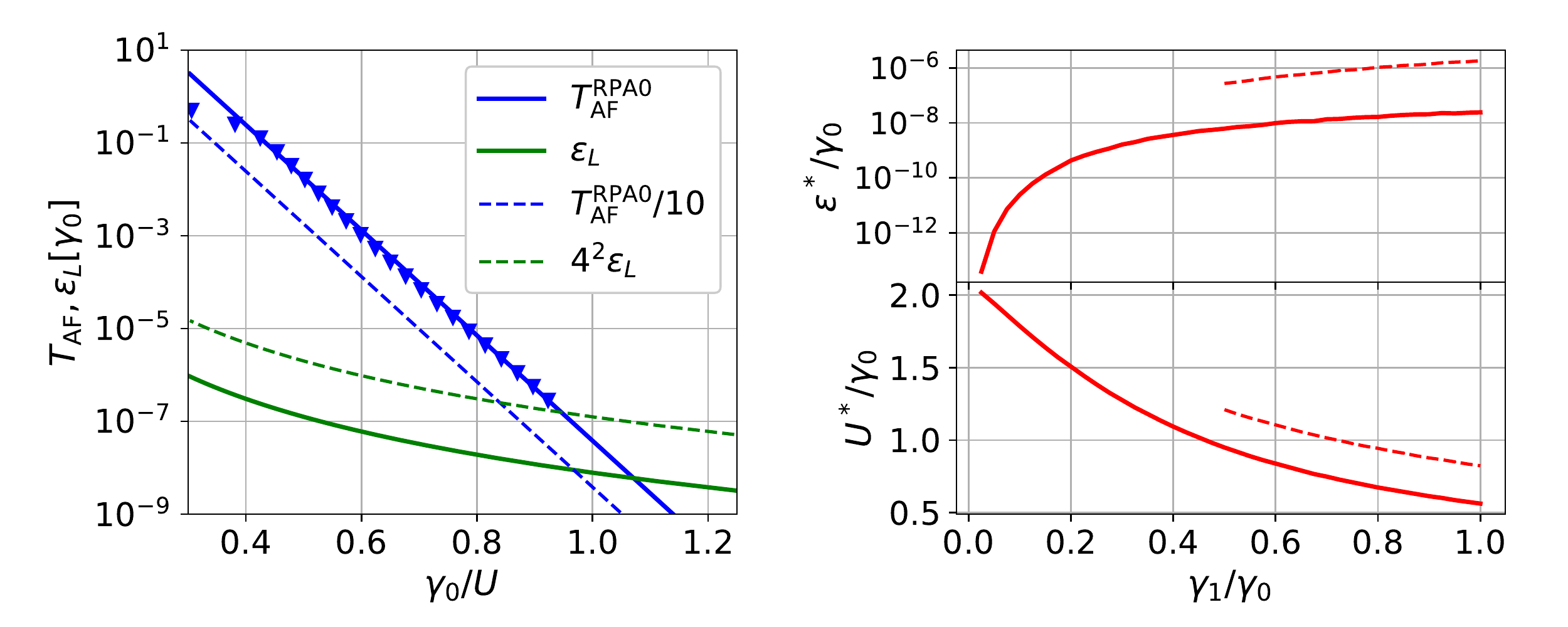}%
 \caption{Left plot: The triangles show the decrease of the AF ordering temperature $T_{AF}$ with $1/U$ in RPA without selfenergy corrections (called RPA0) , following the exponential law (\ref{explaw}) for small enough $U$. The straight solid line is an extrapolation obtained by determining the parameter in (\ref{explaw}) from the data points. The curved solid line is the skew-hopping energy scale $\epsilon_L$ from (\ref{epsL}). The crossing point determines the interaction $U^*$ and the energy scale $\epsilon^*$ below which the skew-hopping term is expected to remove the AF instability. The dashed lines show 'renormalized' behaviors of these tow energy scales, mimicking higher-order and selfenergy effects. Here we increased $\gamma_{43}$ by a 'Stoner' factor of 4 and reduced the AF scale by a factor of 10, as suggested by the self-consistent RPA findings.  Right panel: Dependences of $\epsilon^*$ and $U^*$ obtained form the crossing points as in the left panel on the interlayer hopping $\gamma_1$. The dashed curves show again the (by-hand) renormalized values. }
  \label{TcEpsilonLAna}
\end{figure}

\begin{figure}
 \includegraphics[width=\columnwidth]{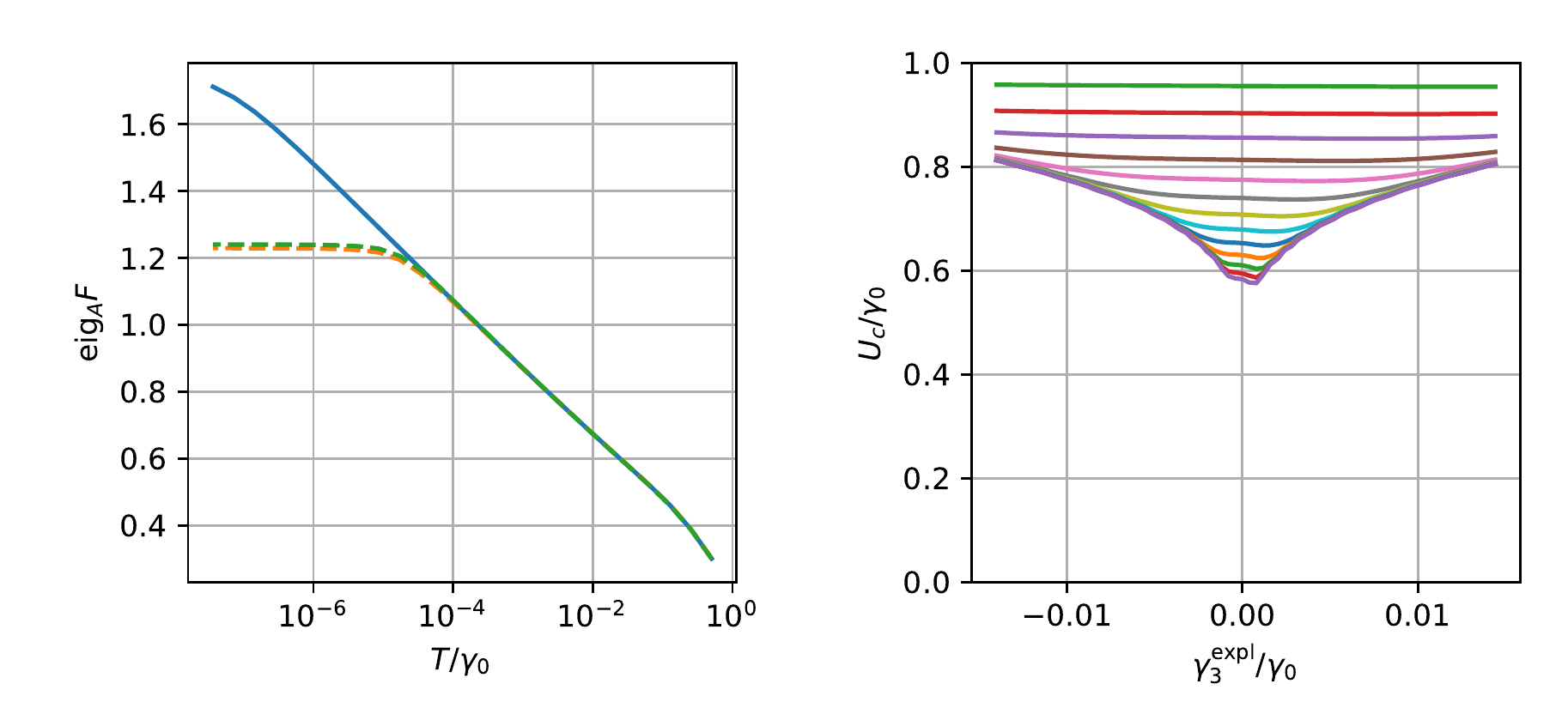}%
 \caption{Left plot: Eigenvalues eig$_{AF}$ corresponding to AF ordering of the RPA0 spin susceptibility (computed with bare propagators) versus temperature $T$. 
 The solid line is for explicit skew hopping amplitude $\gamma_3^{\mathsf{expl}}=0$. It shows a logarithmic increase down to temperatures $\sim 10^{-7}\gamma_0$ below which the numerical resolution seems to become a limiting factor.  
 The dashed lines are for  $\gamma_3^{\mathsf{expl}} \sim \pm 0.013 \gamma_0 $, i.e. in the range of the spontaneously generated values, and cut off  at $T \sim  10^{-5}\gamma_0 \sim \epsilon_L$. Right plot: $U_c = ^/\mathsf{eig}_{AF} $ for temperatures $0.5^n$ with $n$ ranging from 1 (uppermost line) to 25 (lowest line). Near $\gamma_3^{\mathsf{expl}}=$, $U_c$ dips toward zero upon lowering $T$, while it saturates to a nonzero value for $\gamma_3^{\mathsf{expl}} $ away from zero.  }
  \label{chiga3}
\end{figure}

Before going to the self-consistent treatment let us discuss the interaction-induced skew hopping term $\gamma_3$ taken as $\gamma_{43} (\pi T)$ arising from $\Sigma_{43}$ and  $\Sigma_{34}$. In the recent work by Pujari et al.\cite{lang2016} it was suggested that this term might wipe out the AF instability by splitting the quadratic band crossing points into Dirac cones. The energy scale for a relevant change in the dispersion for nonzero $\gamma_3$ is given by the Lifshitz energy\cite{mccann_rev} $\epsilon_L = \gamma_1 (\gamma_3/\gamma_0)^2 /4$.
 With our numerical finding $\gamma_3 \sim 0.001 U^2 \gamma_1^2 / \gamma_0^3$, we get 
\begin{equation}
\label{epsL}
\epsilon_L  \sim \frac{1}{4} \cdot 10^{-6} \cdot \frac{U^4 \gamma_1^5}{\gamma_0^8}  \sim  \frac{(U/\gamma_0) ^4}{4} \cdot 10^{-6} \gamma_0 \quad \mbox{for} \quad \gamma_1=\gamma_0 \, .
\end{equation}
This rather low energy scale competes with the exponentially small energy or temperature scale of the AF instability, 
\begin{equation}
\label{explaw}
\epsilon_{\mathrm{AF}} \sim  W^* e^{-1/\rho_0 U}
\end{equation}
Here we wrote an effective band width $W^*$ which is larger the real bare bandwidth (as the density of states at the Fermi level is much  lower than the average density of states). We now use this relation to extrapolate our numerical data for RPA0 ordering temperature without selfenergies as shown with triangles on the left side of Fig. \ref{TcEpsilonLAna}. The obtained extrapolation line crosses the line for $\epsilon_L$ at a certain $1/U^*(\gamma_1)$.  The so-obtained $U^*$-values are shown in the lower plot on the right side of Fig. \ref{TcEpsilonLAna}.  We find a dominance of $\epsilon_L$ over $ \epsilon_{\mathrm{AF}} $ below $U \sim 0.55 \gamma_0$ for $\gamma_1=\gamma_0$ and  $U \sim 0.93 \gamma_0$ for $\gamma_1=\gamma_0/2$.  
The upshot from these considerations is that - at least in second order in $U$ - the weak-$U$ ground state is indeed a Dirac liquid without magnetic long-range order.  In the left plot of Fig. \ref{chiga3} we show the antiferromagnetic eigenvalue of the RPA spin susceptibility as a function of $T$ without selfenergy corrections but with explicit $\gamma_3^{\mathsf{expl}}$. The solid line for $\gamma_3^{\mathsf{expl}}=0$ shows a logarithmic increase down to temperatures $\sim 10^{-7}\gamma_0$ below which the numerical resolution seems to become a limiting factor.  The dashed lines are for  $\gamma_3^{\mathsf{expl}}\sim \pm 0.013 \gamma_0 $, i.e. in the range of the spontaneously generated values, and cut off  at $T \sim  10^{-5}\gamma_0 \sim \epsilon_L$.  Inverting these eigenvalues leads to RPA critical interaction strengths $U_c$ for AF ordering.   These are shown in the right plot of Fig. \ref{chiga3} for a number of temperatures. One observes the dip toward 0 for  $\gamma_3^{\mathsf{expl}} \approx 0$ and the saturation at nonzero values of $U_c\gamma_0 \lesssim 1$ away from that. Hence, when a $\gamma_3^{\mathsf{expl}}$ is generated perturbatively such that $\epsilon_L$ becomes comparable to the meanfield AF ordering scale at $U\sim1$, the renormalized susceptibility is indeed weakened so much that the AF order for comparable or smaller $U$ should be absent.  This suggests that loss of ground state order due to a finite $\gamma_3^{\mathsf{expl}}$ does occur indeed.

On the other hand, the energy scale for the Dirac cones to dominate over the AF instability is extremely low (see  upper right plot in Fig. \ref{TcEpsilonLAna}). The rather small lower critical $U \le \gamma_0$ where the AF order is wiped out is possibly too low to be consistent with the numerical observations from Ref. \onlinecite{lang2016} where the loss of order was observed at $U\sim 2.5\gamma_0$ for $\gamma_1=\gamma_0$. Hence, if the interpretation of the QMC data and this threshold value is correct, other effects must be present to reduce the ordering tendencies. 

The second-order analysis done so far may be criticized because it might still miss the enhancement of magnetic fluctuations and hence potentially the selfenergy by higher orders in $U$. Moreover we have not yet looked or self-consistency, i.e. the data up to now was obtained without selfenergy corrections. In order to get a more thorough picture of the competition of selfenergy effects and the AF ordering tendencies we now move to a self-consistent RPA study that contains arbitrarily high orders in $U$.

\section{Selfconsistent RPA}
\begin{figure}
 \includegraphics[width=\columnwidth]{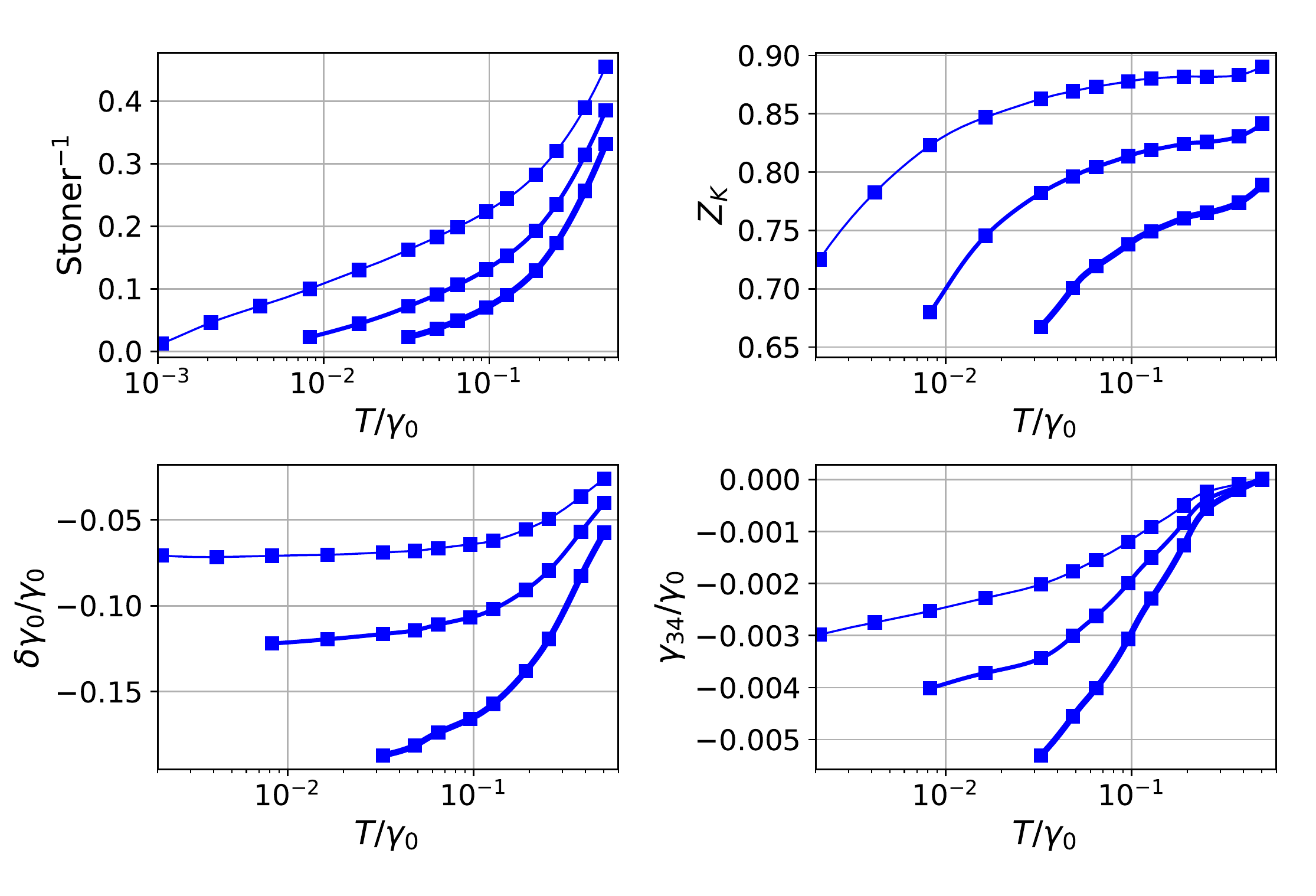}%
 \caption{Selfconsistent RPA selfenergies for different $U$ between $U=2\gamma_0$ (thinnest line) and $U=2.8\gamma_0$ (thickest line). The squares show the data for different temperatures, the lines are interpolations. Left upper plot: Inverse 'Stoner denominator', i.e. $(1-U \mathrm{eig}_{AF})^{-1}$ with the largest eigenvalue $\mathrm{eig}_{AF}$ (see Eq. \ref{chiaf}) of the particle-hole bubble at zero wavevector whose eigenvector is associated with AF ordering. If the 'Stoner denominator' equals zero, the AF instability occurs.  When approaching this situation from higher $T$, the self-consistent numerical procedures runs into convergence problems before zero is reached, hence the lines stop at slightly higher $T$.  Upper right plot: 
Quasiparticle weights $Z_{K,\alpha}=\left[1-i \partial_{i\omega}  \left.  \Sigma_{\alpha\alpha} (\vec{k}, i \omega) \right|_{i\omega=0}\right]^{-1}$ for the orbitals $\alpha=3$ and $\alpha = 4$ (i.e. those that form the bands touching at the Fermi level). Lower left plot: intralayer hopping renormalization at the smallest Matsubara frequency. Lower right plot: Interaction-induced skew hopping at the smallest Matsubara frequency. Data for $\gamma_1=0.5 \gamma_0$. 
}
 \label{TRPAsig}
\end{figure}

For the self-consistent RPA treatment, we compute the selfenergy on a ring around $K$ for a wide range of Matsubara frequencies. While we keep the Matsubara frequency-dependence of the selfenergy explicitly, we use what we learned in the second-order study and exploit the simple wavevector dependences of the selfenergy components. Thus we extract renormalization components $\delta \gamma_{\alpha\beta }(i \omega)$ between the sublattice sites $\alpha$ and $\beta$ in the bilayer unit cell and its nearest neighbors in the same way as described for the second order analysis, but this time with the full ladder RPA interaction (\ref{Vladder}). These are then fed back into the Green's funtions with which the new RPA susceptibility and the new ladder selfenergy are evaluated.  
Various pieces of the so-obtained data are shown in Fig. \ref{TRPAsig}. The upper left plot shows the maximal eigenvalue of the 'Stoner' denominator of Eq. \ref{chiaf} for $U$ ranging between $2\gamma_0$ and $3\gamma_0$ as a function of temperature $T$. Without selfenergy corrections, we would find a linear decrease toward 0 already at rather high $T$, signaling the RPA ordering transition. With the ladder selfenergy contributions fed back self-consistently, the decrease toward 0 is retarded considerably and the lines bend away from 0 when one goes to lower $T$. Therefore the build-up of long-ranged AF correlations toward low $T$ is suppressed significantly. The other panels of the Fig. \ref{TRPAsig} show other components of the self-energy like the $Z$-factors obtained from the frequency-slope of the  site-diagonal selfenergies, the nearest-neighbor hoping renormalization at the smallest Matsubara frequency and interaction-generated skew hopping term $\gamma_{34}(i\omega) $. All show a decrease or growth towards low $T$ but neither $Z$ gest very small toward the AF ordering nor do the hopping renormalizations diverge. Comparing the numbers obtained for $\delta \gamma_{41}$ and $\gamma_{43}$ from the RPA to the second order results, we find a finite but visible 'Stoner enhancement' for these quantities of about a factor $\sim 4$.

\begin{figure}
 \includegraphics[width=\columnwidth]{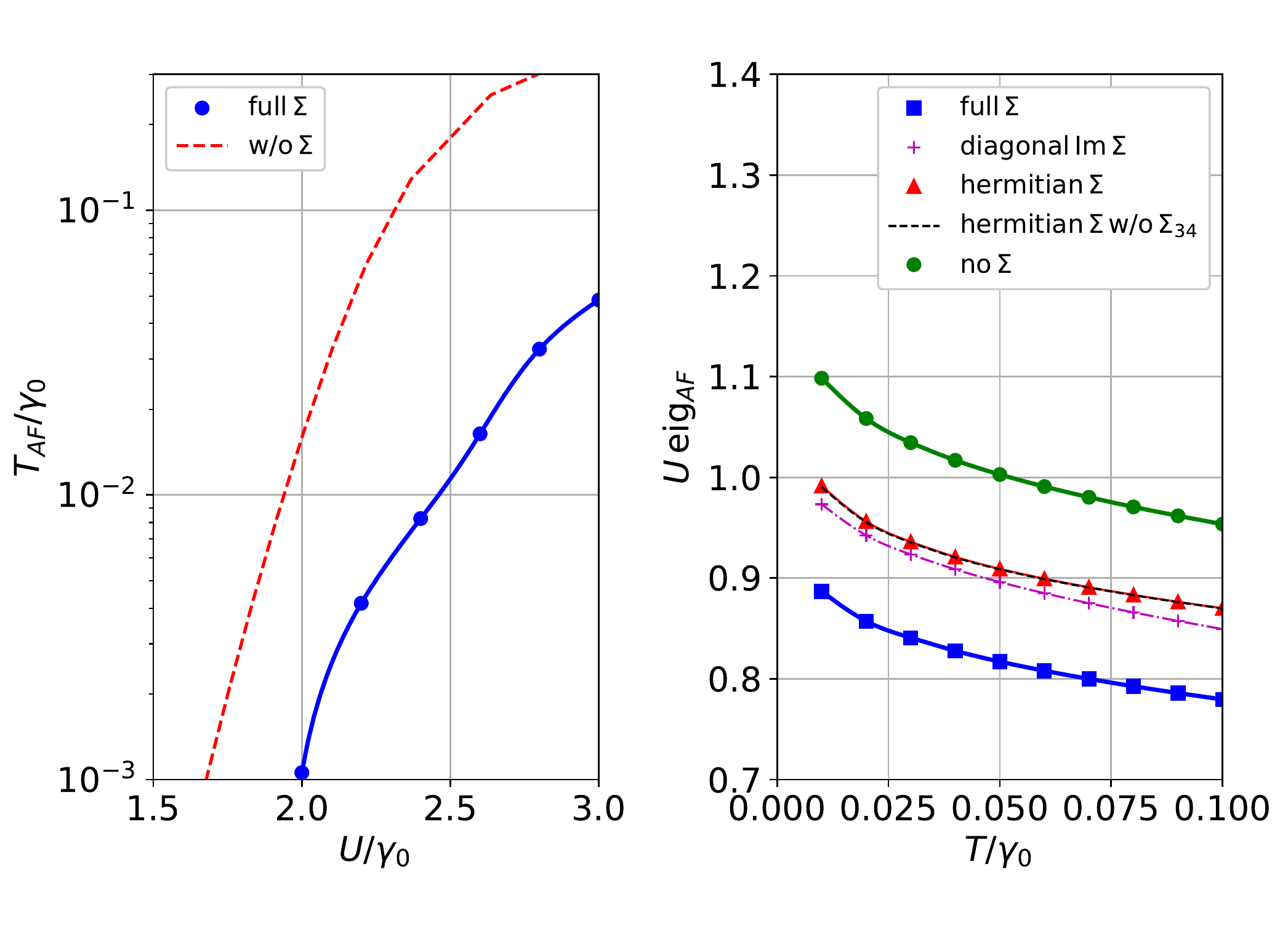}%
 \caption{Left plot: Data for $\gamma_1=0.5 \gamma_0$. The bullets show the temperatures where $U \mathrm{eig}_{AF} $ reaches 0.9875 in self-consistent RPA, i.e. give an estimate of the AF (short-range) ordering temperature including selfenergy effects.  The dashed line is the RPA0  AF ordering scale without selfenergy corrections, which is more than one order of magnitude higher. Right: $U \mathrm{eig}_{AF} $ with renormalized propagators with different self-energy parts switched on. The lowest line is with full self-energy corrections, the upper-most line without selfenergy. The two lines in the middle are with only intralayer hopping renormalization and with only quasiparticle degradation (diagonal selfenergy). 
 Data for $\gamma_1=0.4 \gamma_0$, $U=2.2\gamma_0$. }
  \label{TcsSC}
\end{figure}

This data shows clearly that the combined impact of the selfenergy components impedes the AF instability strongly. In fact it is not clear from this data that the AF instability still survives for larger $U$, but deciding whether the transition still survives for low enough $T$ is probably a question that anyway cannot be answered in a controlled way from RPA. Here we are satisfied to observe the significant reduction of the ordering tendencies. In Fig. \ref{TcsSC} we plot a line where the 'Stoner denominator' from the previous figure gets less than 0.0125, i.e. the system gets close to the instability. Getting even closer to the pole is numerically challenging because of a rather unstable convergence behavior of the self-consistency cycle close to the instability. Hence these data points represent an upper estimate of ordering temperatures $T_{AF}$.   On the logarithmic plot we observe that the ordering temperatures $T_{AF}$ have been suppressed by more than one order of magnitude compared to the case without selfenergies. 

We can now ask if the RPA data with self-consistent selfenergy feedback changes the conclusions we found regarding the competition between the skew hopping selfenergy and the AF instability. While the AF ordering $T_c$ is lowered strongly by the selfenergy, the skew-hopping term is 'Stoner'-increased by roughly a factor of four in RPA. Based on this and and a suppression of the AF ordering temperature by one order of magnitude, we can redo the analysis of Fig. \ref{TcEpsilonLAna} with renormalized quantities. If we use the (by-hand) renormalized values (see dashed lines) with a Stoner enhancement of 4 and a $T_{AF}$-reduction of 10, the $U^*$- and $T^*$- values do not change drastically. We do not show the behavior for smaller $\gamma_1$ as we did not track the Stoner enhancement for these values. Thus, given the small energy scales, it should be hard to observe the effect that the interaction-induced skew hopping term kills the AF ordering.  Due to our limited numerical resolution we cannot perform self-consistent calculations at these low scales. 

On the other hand, our calculations show that the AF order is significantly affected by other selfenergy effects, mainly by quasiparticle degradation and Dirac cone steepening. This picture is corroborated by computing the AF ordering eigenvalue for different selfenergy components switched on and off. In the right plot of Fig. \ref{TcsSC} we show these eigenvalues as a function of $T$ for $\gamma_1=0.4 \gamma_0$. The 'no $\Sigma$'-line shows the growth of the eigenvalue for lower $T$ without selfenergies included, with a crossing of unity around $T=0.5\gamma_0$. The 'full $\Sigma$'-line shows the slowed-down growth of the eigenvalue for lower $T$  with all components of the self-consistent selfenergy included. The line in between are calculations of the eigenvalue where the certain selfenergy components of the 'full $\Sigma$'-selfconsistent data $\Sigma_{ab}$ were set to zero on the internal lines of the particle-hole loop. Only taking nonzero hermitian selfenergy components, i.e. frequency-dependent hopping renormalizations, into account, we get roughly the same reduction of the eigenvalue as only taking the non-hermitian sublattice-diagonal self-energy terms into account. So, these two selfenergy contributions account each for half of the reduction. If we again take only hermitian nonzero selfenergy terms but in addition leave out the skew-hopping selfenergy, i.e. set $\Sigma_{43}=\Sigma_{34}=0$, the AF eigenvalue does not change in any visible way (by less than one permille). Similarly, the interlayer hopping renormalization does not have any noticeable effect.
Hence the in-plane hopping increase and the quasiparticle degradation are the two decisive players in the reduction of the AF ordering tendencies.

 \section{Discussion}
We have presented a self-consistent RPA analysis of antiferromagnetic (AF) ordering tendencies in the Bernal-stacked bilayer Hubbard model on the honeycomb lattice. Compared to the AF transition temperatures obtained meanfield theory or in RPA without selfenergy corrections, the selfenergy-corrected transition temperatures are lowered by more than an order of magnitude. This definitely shows that selfenergy corrections are important to consider at least if quantitative questions about energy scales and temperatures are posed. Ordering tendencies that appear readily at high scales in such band-crossing point models may be more fragile than e.g. found in the Hubbard on the square lattice\cite{uebelacker}. The identified the upward-renormalization of the in-plane hopping and the quasiparticle degradation by the frequency-dependence of the sublattice-diagonal selfenergy have been shown to be the two main factors for the reduction of the AF ordering tendencies.

Of course, this RPA analysis presented here is still not fully controlled or exact. There are are other effects that influence a potential AF ordering. From renormalization group  studies it is well known that particle-particle diagrams not included in the RPA lead to a further reduction of the ordering tendencies. For instance, in the single-layer honeycomb Hubbard model, the non-selfconsistent RPA critical interaction strength for AF ordering is $U_c^{MF} \sim 2.2\gamma_0$, while the functional RG including the particle-particle channel but no selfenergies gives $U_c^{fRG} \sim 2.8\gamma_0$\cite{sanchez}. This is almost half way to the QMC value $U_c^{QMC} \sim 3.6\gamma_0$. One may speculate that including selfenergy corrections on top of this explains the remaining difference.  Our study of the selfenergy effects within RPA that shows a strong effect and thus provides some hope that this may indeed work. Exploring this possibility in a simpler RPA framework was one of the motivations for this work.  
 
On the other hand, with our numerical approach using discrete Matsubara frequency dependences we hesitate to go to very low temperatures or $T=0$.  At nonzero $T$, in a truly two-dimensional system, AF ordering should not occur due to spin-wave fluctuations. Nevertheless we should expect that the temperature scale $T_{AF}$ worked out here where the AF spin susceptibility becomes large, i.e. where the RPA transition takes place in our study, is still a relevant emergent energy scale of the system that, e.g., should be observable by significant short range order and the opening  of an energy gap in the fermionic spectrum whose size approaches a value comparable with $T_{AF}$ at $T=0$. Therefore we expect that $T_{AF}$ is a good measure of the strength of AF ordering in the system. It would actually be very interesting to see if e.g. finite-$T$ QMC calculations confirm these reduced temperature scales and the suggested spectral reconstructions at $T_{AF}$.

We have also investigated the question if an interaction-induced skew-hopping term can wipe out the AF ordering tendencies by removing the quadratic band crossing points. We computed this term in second order perturbation theory and in RPA and determined its magnitude and sign. Based on our data, the removal of the quadratic band crossing points by the formation of four Dirac points should indeed happen, albeit at rather low U $\lesssim \gamma_0$ for  interlayer hopping $\gamma_1 $. The energy scales at which these spectral changes matter are extremely small. Thus we have doubts  that our analysis explains quantitatively the QMC data of a recent publication\cite{lang2016}, suggesting a loss of AF order already at intermediate $U$-values two or three times larger than found from our skew-hopping term analysis. Naively, one might speculate that the other mentioned selfenergy terms that are much bigger and reduce the AF energy scales by more than one order of magnitude  cause a disappearance of the AF order for finite-size systems already at intermediate $U$-values.   
  
CH thanks the German Research Foundation for support, in particular through the priority program SPP1459 on graphene and the research training group RTG1995 on quantum many-body methods. We acknowledge useful discussions with Th. Lang, Jie Yuan, S. Wessel, Z.Y. Meng, Q.H. Wang, R. Kaul, L. Janssen, O. Vafek and M. Scherer.

\end{document}